
\magnification=1200
\def\ni{\noindent}
\def\.{\mathaccent 95}
\def\a{\alpha}

\def\ga{\gamma}

\def\ep{\epsilon}

\def\ka{\kappa}
\def\la{\lambda}

\def\Ga{\Gamma}

\def\frac#1#2{{\textstyle{{#1}\over {#2}}}}
\def\ni{\noindent}
\def\lsim{\mathrel{\rlap{\lower4pt\hbox{\hskip1pt$\sim$}}
    \raise1pt\hbox{$<$}}}
\def\gsim{\mathrel{\rlap{\lower4pt\hbox{\hskip1pt$\sim$}}
    \raise1pt\hbox{$>$}}}
\def\sqr#1#2{{\vcenter{\vbox{\hrule height.#2pt
         \hbox{\vrule width.#2pt height#1pt \kern#1pt
         \vrule width.#2pt}
         \hrule height.#2pt}}}}

\newbox\grsign \setbox\grsign=\hbox{$>$} \newdimen\grdimen \grdimen=\ht\grsign
\newbox\simlessbox \newbox\simgreatbox
\setbox\simgreatbox=\hbox{\raise.5ex\hbox{$>$}\llap
     {\lower.5ex\hbox{$\sim$}}}\ht1=\grdimen\dp1=0pt
\setbox\simlessbox=\hbox{\raise.5ex\hbox{$<$}\llap
     {\lower.5ex\hbox{$\sim$}}}\ht2=\grdimen\dp2=0pt

%
%

\def\ref#1  {\noindent \hangindent=24.0pt \hangafter=1 {#1} \par}

\def\doublespace {\smallskipamount=6pt plus2pt minus2pt
                  \medskipamount=12pt plus4pt minus4pt
                  \bigskipamount=24pt plus8pt minus8pt
                  \normalbaselineskip=24pt plus0pt minus0pt
                  \normallineskip=2pt
                  \normallineskiplimit=0pt
                  \jot=6pt
                  {\def\smallskip {\vskip\smallskipamount}}
                  {\def\medskip   {\vskip\medskipamount}}
                  {\def\bigskip   {\vskip\bigskipamount}}
                  {\setbox\strutbox=\hbox{\vrule 
                    height17.0pt depth7.0pt width 0pt}}
                  \parskip 12.0pt
                  \normalbaselines}


%

\def\ts{\times}

\def\ts{\times}

\centerline{\bf DISTINGUISHING SOLAR FLARE TYPES}
\centerline{\bf BY DIFFERENCES IN RECONNECTION REGIONS}
\medskip
\centerline{Eric G. Blackman}
\medskip
\centerline{Institute of Astronomy, Madingley Road, Cambridge CB3 OHA, England}
\medskip
\centerline{Email: blackman@ast.cam.ac.uk}
\centerline{(accepted to ApJ Lett.)}
\medskip
\doublespace
\centerline{\bf ABSTRACT}

Observations show that magnetic reconnection and its slow shocks occur
in solar flares.  The basic magnetic structures are 
similar for long duration event (LDE) flares 
and faster compact impulsive (CI) flares, but the former require less 
non-thermal electrons than the latter. 
Slow shocks can produce the required 
non-thermal electron spectrum for CI flares by Fermi acceleration
if electrons are injected with large enough energies to resonate 
with scattering waves.  The dissipation region may provide the injection
electrons, so the overall number of non-thermal electrons reaching the
footpoints would depend on the size of the dissipation region 
and its distance from the chromosphere.
In this picture, the LDE flares have converging inflows toward a 
dissipation region that spans a 
smaller overall length fraction than for CI flares.  
Bright loop-top X-ray  spots in some CI flares can be attributed to 
particle trapping at fast shocks in the downstream flow, the presence of which
is determined by the angle of the inflow field and velocity to the slow shocks.
  
\noindent ${\bf Subject\ Headings:}$ Magnetic Fields: MHD; 
Acceleration of Particles; Sun: magnetic fields, flares, corona.

\vfill
\eject

\centerline{\bf 1. Introduction}

Solar flares are rapid bursts of radiation from the solar atmosphere. 
They have long been believed to result from the conversion of magnetic energy 
(e.g. Pneuman 1981) by magnetic reconnection
(e.g. Biskamp 1994), a process in which
oppositely magnetized flows merge across a thin 
dissipation region (DR).  Recent flare X-ray observations from 
the Yohkoh satellite (e.g. Tsuneta et al. 1992; Masuda et al. 1994) have 
confirmed that  reconnection is fundamental to the observed energy release.

Flares are divided into two classes (Pallavicini 1991),  
the long duration event (LDE) two-ribbon flares
and the compact impulsive (CI) flares.  The former,
with  typical durations of hours, and luminosities
$\sim 10^{28}$erg/sec,  
have been modeled as the merging of
magnetic field lines at the top of an inverse-Y 
type field line configuration (Fig. 1).  Downward
ejected plasma heats photospheric footpoints, 
inducing a flux loop filling upward flow that generates
soft X-rays (e.g. Sturrock 1966; Tsuneta 1996ab). 
LDEs are of order $10^4-10^5$ km in height, have generally smooth time
profiles and not much non-thermal electron emission (e.g. Tsuneta 1996a).  

CI flares are $\lsim 1/10$ the size of LDE flares, lasting of order minutes,
and with similar luminosities (e.g. Masuda et al. 1994).
They show strong impulsive phases with bursts of non-thermal
emission and variability times $\sim {\rm O}(0.1)$ sec.
One of the intriguing implications of the CI Masuda flare 
(Masuda et al. 1994) observations
is that the fundamental inverse-Y configuration
and the downward plasma flow are likely common to the CI as well 
as to the LDE flares, in contrast to what was previously thought.  
The CI Masuda flare is also interesting for its
hard X-ray source at the top of its soft X-ray loop in addition to
the usual hard X-ray footpoints.  

Yohkoh observations indicate the presence of slow shocks 
in at least some flares (Tsuneta 1996ab).  
This provides support for Petschek (Petschek 1964) or Sonnerup (Sonnerup 1970)
type rapid reconnection models, i.e. those
in which slow shocks extend from the corners of the thin DR,  
dividing the inflow and outflow.  
The length of the slow shocks vs. the length of the
DR likely depends on boundary and inflow conditions
(Priest \& Forbes 1986; Forbes \& Priest 1987).  
Priest \& Forbes (1986) have shown that a variety
of solutions can be obtained by 
changing the angle of inflow velocity to the reconnection
region.

Though the basic reconnection configuration and the  
plasma filling of a soft X-ray loop are common to LDE and CI flares,
the relative sizes of the DR and its shocks and their 
distance from the chromosphere are likely important in determining 
differences in the number of non-thermal electrons reaching the footpoints.
Section 2 addresses how reconnection
slow shocks may be a source of particle acceleration.
Section 3 describes how the 
DR is important in injecting electrons
into the shocks and determining 
the extent of non-thermal acceleration.  
In section 4, the condition for a downstream fast shock is derived 
in terms of the inflow parameters. 
In section 5, a more specific discussion distinguishing flares is given, and
section 6 is the conclusion.



\centerline{\bf 2. Acceleration at Slow Shocks}

Flare reconnection occurs at the very top
of the configuration of Fig. 1 
as regions of oppositely magnetized plasma flow in from the 
sides and intersect at the thin DR.
The magnetic annihilation produces a topology change with an 
X-point at the interface.  The shocks occur at the boundaries
between inflow and outflow.
Unlike fast shocks, slow shocks have a
weaker magnetic field downstream than 
upstream. Fermi acceleration is still possible at these 
shocks (Blackman \& Field 1994; Blackman 1996).  To see this, 
note that the MHD shock jump conditions for mass, momentum, energy, and electromagnetic
fields are (e.g. Melrose 1986)
$$\rho_1v_{1n}=\rho_2v_{2n},\eqno (1)$$
$$\rho_1 v_{1n}^2+P_1+B_{1t}^2 / 8 \pi=\rho_2 v_{2n}^2+P_2+B_{2t}^2/8\pi\ ;\ 
\rho_1v_{1n}{\bf v_{1t}}-B_{1n}{\bf B_{1t}}/4\pi
=\rho_2v_{2n}{\bf v_{2t}}-B_{2n}{\bf B_{2t}}/4\pi,\eqno(2)$$
$$(1/2)\rho_1v_1^2v_{1n}+\Gamma(\Gamma-1)^{-1}P_1v_{1n}+(B_1^2/4\pi)v_{1n}-{\bf v_1\cdot B_1}B_{1n}/4\pi
=(1/2)\rho_2v_2^2v_{2n}$$
$$+\Gamma(\Gamma-1)^{-1}P_2v_{2n}+(B_2^2/4\pi)v_{2n}-{\bf v_2\cdot B_2}B_{2n}/4\pi,\eqno(3)$$
$$ B_{1n}=B_{2n}\ ;\ ({\bf v_1}{\rm x}{\bf B_1})=({\bf v_2}{\rm x}{\bf B_2}).\eqno(4)$$ 
where $B$ is the magnetic field, $v$ is the velocity, 
$P$ is the pressure, $\rho$ is the density and $\Gamma$ is the adiabatic
index.  The subscript $1(2)$ refers to the up(down)stream
region, and the subscript $n(t)$ refers to the normal (tangential)
components.  

The shock is $\perp$ to the ${\hat {\bf n}}, {\hat {\bf y}}$
plane as shown in Fig. 2.  We assume the switch-off condition,
$B_{2y}=0$, and also that ${\bf v}_1/|v_1|\cdot {\hat{\bf y}}<< 1$. 
Define ${\tilde C}\equiv cos\theta$, ${\tilde S} \equiv sin\theta$ 
and ${\tilde T}\equiv tan\theta$ where $\theta$ is the 
angle between the downstream flow and the shock normal.
Define $C_{1} \equiv cos {\phi}_{1}$, $S_{1}
\equiv sin {\phi}_{1}$ 
and $T_{1} \equiv tan { \phi}_{1}$  where $  \phi_{1}$ 
is the angle between the upstream field and the shock normal.
The configuration of Fig. 2 is then described by
$$v_{1n}=-v_1 ,\ B_{1n}=-B_1 C_1,\ B_{1y}=-B_1 S_1, 
v_{2n}=-v_2 {\tilde C},\ v_{2y}= v_2 {\tilde S},\ B_{2n}=-B_2,\ v_{1y}=B_{2y}=0. \eqno(5)$$
For $\Ga=5/3$ and $\beta_1 \equiv a_{1s}^2/v_{1A}^2 <<1$, where 
$a_{1s}$ and $v_{1A}$ are the inflow sound
and Alfv\'en speed, plugging (5) into (1-4) gives
$$T_1^2=2(r_s-1)(r_s-4)/(5r_s-2r_s^2),\eqno(6)$$
$$\beta_2=(5/3)[(r_s-1)/r_s+T_1^2/2]\eqno(7)$$
$$M_{2A}^2\equiv v_2^2/v_{2A}^2=(1+r_s^2T_1^2)/r_s,\eqno(8)$$
where $M_{2A}$ is the outflow Mach number, $v_{2A}$ is the outflow
Alfv\'en speed, $\beta_2\equiv a_{2s}^2/v_{2A}^2$, $a_{2s}$ is the outflow
sound speed and $r_s\equiv\rho_2/\rho_1$.
Since $T_1^2>0$, 
$(8)$ shows that $2.5\le r_s\le 4$ for a low 
$\beta_1$ switch-off shock (Kantrowitz \& Petschek 1966;  Blackman
\& Field 1995), with the lower limit being a
perpendicular $(\perp)$ shock and the upper limit a parallel $(||)$ shock.

The equation for diffusion and convection of particles 
across a shock is given by (Jones \& Ellison 1991)
$$\partial_n[v_n f-\ka_n \partial_n f]-
(1/3)(\partial_n v_n)\partial_{p}[p f]=0,\eqno(9)$$
where $f$ is the particle distribution function, 
$v_n$ is the normal flow velocity across the shock, 
 $p$ is the particle momentum, $\ka_n\sim p\la/3m$ 
is the normal diffusion coefficient,  $\la$ is the mean-free path between
particle-wave scatterings, and $m$ is the particle mass.
Fermi acceleration operates as particles diffuse between 
scattering centers (presumably MHD Alfv\'en turbulence) on each side of the 
shock.  Particles always see the centers converging, as the normal velocity is
larger upstream.  The solution of $(9)$
across the shock with thickness $<$ mean-free path (Jones \& Ellison 1991)
shows that the outflow particle 
spectrum for a steeper inflow spectrum takes the
power law form $f\propto p^{-\a}$, with index 
$\a=(r_s+2)/(r_s-1)$, 
where $p$ is related to the energy
$E$ by $p=E^{1/2}(E+2mc^2)^{1/2}/mc$. 
Thus for weakly or non-relativistic particles, $f(E)\propto E^{-2\a}$.    
For $2.5<r_s<4$, $ 4 \le 2\a \le  6$ for the slow shocks, which is
quite consistent with the required electron spectra
derived by inverting the observed photon spectrum from
thick target  models of X-ray footpoints of solar flares
(Aschwanden \& Schwartz 1996).
Slow shocks may therefore supply non-thermal electrons.

Though $4 <2\a< 6 $  results from an analytical 
treatment, shock acceleration is a very non-linear process.
Fermi acceleration can be even more efficient
in the non-linear regime, transferring $\ge 1/2$ of the inflow energy to particles (Jones \& Ellison 1991). Similar beam  instabilities to those which
signature Fermi acceleration in fast shocks have also been seen in 
slow shock simulations (Omidi \& Winske 1994).
Support for slow shock acceleration 
is present in the geomagnetic tail where
turbulence, required for Fermi acceleration, is seen on both sides of the
shock fronts (Coroniti, et al. 1994) and
non-thermal tails in the electron spectra are seen (Feldman et al. 1990).

\centerline{\bf 3. Electron Injection and Solar Flare Reconnection}

For CI flares, unlike LDE flares,
$>20$keV non-thermal electrons may contribute to of 
order the total luminosity (Lin \& Hudson,  1971). 
Electrons can only be shock accelerated when they are injected
above a critical energy a factor of the  mass ratio $(m_p/m_e)$ 
higher than that required by protons. 
Fermi acceleration requires downstream particles to scatter upstream and
gain energy from turbulent scattering centers that boost only
the momentum parallel to the magnetic field.  Pitch-angle randomizing 
must occur in order for multiple energy gains to be imparted 
(Eilek \& Hughes 1991).  
This randomization is provided by resonant Alfv\'en waves which 
exist with frequencies only below the ion gyro-frequency.
 
For Alfv\'en turbulence (Eilek \& Hughes 1991)
in the limit that the Alfv\'en speed exceeds $m_e c/m_p$, 
the lower bound on the Lorentz factor for electrons 
to resonate with Alfv\'en waves is 
$\ga_e \gsim 1+(m_p/m_e)(v_{2A}/c)^2$
for $ v^2_{2A} < (m_e/m_p)^2 c^2$ while for $v^2_{2A}>(m_e/m_p)^2c^2$, 
$\ga_e\gsim (m_p/m_e)(v_{2A}/c)$.  
For reconnection shocks, stochastic acceleration (e.g. Larosa 1996)
in the DR may provide injection electrons 
with self-generated Alfv\'en waves.  To see 
that this is kinematically feasible,   
note that upon absorbing the annihilated field energy in the DR, 
the average $\gamma_e$ there could be $\sim 1+(v_{1A}^2/c^2)(m_p/m_e)$.
Since $v_{1A}\ge v_{2A}$ for a slow shock, 
the DR can in principle always inject.
Since all field lines in a reconnection region which
pass through the shock also pass through the DR, 
the fraction of electrons  that could be
injected and accelerated is at least the fraction that passes through
the DR.  

\centerline{\bf 4. Formation of and Acceleration at Fast Downstream Shocks}

When ${\bf B}_2\cdot {\bf v}_2/|B_2v_2|<< 1$, $v_2$
will be supermagnetosonic (Melrose 1986) when
$v_{2}^2=v_{2n}^2+v_{2y}^2>a_{2s}^2+v_{2A}^2$.
Using (1-5) this reduces  (Blackman \& Field 1994)
to $6r_s^2-13r_s-20 < 0,$
which is satisfied for 
$r_s \lsim 3.2$ or $T_1>1.25$ from  $(6)$.
A supermagnetosonic outflow becomes the condition for
a fast shock when the field is line-tied at the outflow boundary.  
The jump conditions (1-4) across such a
quasi-$\perp$ fast shock for $\Ga=5/3$ give  
$$M_{2A}^2=3r_f\beta_2/(4-r_f)+(3/2)r_f(r_f-1)/(4-r_f),\eqno(10)$$
where $r_f$ is the compression ratio across the fast shock.
Using (6-8) and (10), it can be shown that $1 \lsim r_f \lsim 2$ when $2.5\lsim r_s\lsim 3.2$.
The inverse dependence is expected because a decrease in $r_s$
 corresponds to an increase in tension force along the
shock plane, and thus a larger $M_{2A}$, accounting for the
larger $r_f$.  

Thus, the stronger the inflow field tangential to the slow shock, 
the more likely the presence of
an outflow fast shock.  But the condition
for a fast shock was determined in the frame for which 
the inflow is $\perp$  to the shock.  
If the lab frame inflow has a tangential component  
parallel to that of the outflow, the minimum $T_1$ for a shock would
decrease, while a tangential component
opposite to the outflow would increase the minimum $T_1$.
The absence or presence of a downstream fast shock can be used
as a diagnostic to determine whether the flow is converging or diverging
on each side of the DR.  This determines the mode of reconnection 
(Forbes \& Priest 1986).

\centerline{\bf 5. Application to Flares}

The bright X-ray source above the CI Masuda flare loop top 
may be associated with favorable inflow conditions for an
outflow fast shock.  Time of flight (TOF) analyses (Aschwanden et al.  1996)
require loop-top electron trapping
either by collisional trapping, or from an enhanced site
of Alfv\'en waves.  The presence of a fast 
shock as a site of turbulence
may provide the latter.  Since $r_f<r_s$ as computed above, 
trapping, rather than acceleration, could be the fast shock's primary purpose.
The TOF analysis of the Masuda flare
also suggests that the actual acceleration region is located above the 
loop-top X-ray source (Aschwanden et al. 1996) which is consistent with the present picture.
Also consistent is the fact that the loop top source and the
footpoint sources are observed to mimic each others' temporal behavior.
If slow shocks are responsible for acceleration, some of the fast particles
will escape toward the loop top, and others toward the footpoints, 
with the source of acceleration being the same for both.

In the approach of Larosa et al. (1996), stochastic turbulence
in the outflow of a reconnection region is suggested as a
possible source of CI flare electron acceleration. 
It is likely that stochastic and slow shock acceleration 
have a symbiotic relationship. Stochastic acceleration near the
DR could provide the injection electrons which are
subsequently accelerated along the shock.  
Away from the DR in the outflow, the field
is very small, and the stochastic acceleration would not be
effective there.  It would be difficult to explain the differences between
loop top X-ray source CI flares and those which just have
brightened footpoints without at least invoking a downstream fast
shock.  But the canonical compression ratio of the fast shock, as estimated
above, may not be high enough to account for all of the acceleration.
Since it is known that in LDE flares the slow shocks are involved in heating
(Tsuneta 1996a), their role for electron acceleration in CI flares may also 
be important as described herein.

Stochastic acceleration provides the upper limit on the variability
time scale, $t_{sh}$, resulting from injection+shock Fermi acceleration, since 
all turbulent scattering collisions in the latter mechanism  
`head-on' making it more efficient.  
This gives $t_{sh}\sim \ka_n/v_1^2 \lsim  t_{sto} \sim w/v_{1A}$ (Larosa et al. 1996) 
where $w$ is the width of the reconnection outflow.  For the
Masuda flare, $v_{1A}\sim 5\ts 10^8$cm/sec and $w\lsim 10^8$cm
so that $t_{sh}\lsim 0.25$sec,  consistent with time scales of
CI flare spikes (e.g. Aschwanden et al. 1995).
Later we find the thermalization time of electrons is larger than $t_{sto}$,
so that Fermi acceleration should dominate all Coulomb collisions
at the acceleration sites.



For non-thermal electrons to reach the footpoints, 
the electrons must not collisionally thermalize before arriving.  
The time scale for a density of electrons, $n$, with average energy $\ep$ to 
thermalize is given by (Stepney 1983)
$t_{th} \sim 8.5 (\epsilon/25{\rm keV})^{3/2}(n/10^{10}{\rm cm^{-3}})^{-1}
({\rm ln}  \Lambda/20)^{-1}$sec, 
where ${\rm ln}\Lambda$ is the Coulomb logarithm.
Avoiding thermalization requires that the distance from the shocks to the
footpoints at least satisfies $D < [\xi (2\ep/m_e)^{1/2}+(1-\xi )v_2]
t_{th}$, where $\xi$ is the fraction of accelerated particles that
can move directly along a field line to a footpoint. 
(Many of the accelerated particles will come from the outflow region
where the field is horizontal, and
because of the their small gyro-radii even in the outflow, they
can only convect to footpoints at the outflow speed.)
The above  condition on $D$ might not be met
for an LDE flare with small $\xi$
since the outflow velocity (e.g. Tsuneta 1996a)
is $v_2\sim v_{1A}/r_s^{1/2} \sim 10^8$cm/sec.  Thus $v_2t_{th}\sim 1.4 \ts 10^9$cm, which is too small by an order of magnitude or so. 
However, for the typically smaller CI flares, the reconnection outflow is  
$\gsim 5$ times faster 
(e.g. Masuda et al. 1994) so $v_2t_{th}\sim 7\ts 10^{9}$cm, and this would be
large enough to allow canonical non-thermal electrons to convect to typical
CI flare footpoints (Aschwanden et al. 1996).

The number of electrons/sec, $N$, reaching the
footpoint sites for CI flares
 would be comprised of those injected first
by the DR.  This is given by
$N \sim t_{sh}^{-1}n V_{dif}$, where the DR volume is
$V_{dif}\sim R^2 h\sim R^3 (v_{1}/r_s v_{2})$ 
with $R$ and $h$ the DR length and thickness (Fig. 2), 
and the last similarity follows from mass flux conservation through the DR.
Now $v_1 \sim 8\pi L_{nt}/B_1^2 R^2$ where $L_{nt}$ is the non-thermal 
luminosity and $v_2\sim v_{1A}/r_s^{1/2}$.
Thus 

\noindent $N \sim 10^{35}(t_{sh}/0.25 {\rm sec})^{-1}(n/ 10^{10}{\rm cm^{-3}})^{3/2}
(B_{1}/200{\rm G})^{-3}(L_{nt}/10^{28}{\rm erg/sec}) 
(r_s/ 3)^{-1/2}(R/4 \ts 10^{8}{\rm cm}) {\rm sec}^{-1}$.
The scalings are reasonable for the CI Masuda flare.  
Electron injection from the DR 
followed by further processing in the shocks is therefore feasible.
Note that $R$ need only be $\sim 1/15$ of the overall height to the DR
in the Masuda flare.  

 

For LDEs the DR length fraction would be even less.
The DR of the Feb 21 LDE flare 
(Tsuneta 1996a) occupies  a tiny fraction of the 
overall region, qualitatively consistent with the present picture.
The absence of a fast shock in this LDE suggests
that the inflows to the DR are converging flows.
This is because $T_1 \sim 5$  (Tsuneta 1996a)
which is above the threshold for a downstream fast shock, 
so the inflows must be converging 
to make the effective condition for an outflow shock more stringent.
This is also consistent with having a small DR in the unified 
reconnection models of  Priest \& Forbes (1986).





\centerline{\bf 6. Conclusions}

Different combinations of slow and fast shocks and stochastic acceleration 
may lead to different flare types from similar basic inverse-Y structures.
Non-thermal acceleration from slow shocks is 
likely most effective when the thin DR
is long enough to provide injection electrons
which can subsequently be shock accelerated.
The canonical spectrum from Fermi acceleration at slow reconnection
shocks reasonably matches the required CI flare electron spectrum 
(Aschwanden \& Schwartz 1996).  
If the DR were very large, then stochastic Fermi acceleration could dominate.
To summarize, the absence of significant non-thermal electrons reaching the
footpoints for LDE flares would result from both 
a small DR length and its larger distance
from the footpoints compared to CI flares.  
CI flares with loop top X-ray sources would 
have a very strong inflow magnetic field component along the slow shocks,
enabling an outflow fast shock to form.


Acknowledgements:  Thanks to G.B. Field for sending me Tsuneta (1996b).

\noindent Achterberg, A., 1987, 
in $Astrophysical\ Jets\ and\ Their\ Engines$, W. Kundt ed.,
NATO series C, vol 208, (Dodrecht:  Reidel).

\noindent Aschwanden, M.J., Schwartz, R.A., \& Alt, D.A., 
1995, ApJ, {\bf 447}, 923.

\noindent Aschwanden, M.J. et al., 1996, ApJ, {\bf 464}, 985.

\ni Aschwanden, M.J., \& Schwartz, R.A., 1996 {\bf 464}, 974.




\noindent Biskamp,D., 1994, Phys.  Rep., $\bf 237$, 179.

\noindent Blackman, E.G. \& Field, G.B., 1994, Phys Rev. Lett, $\bf 73$, 3097.

\noindent Blackman, E.G., 1996, ApJ, {\bf 56}, L87.

\noindent Coroniti,F.V. et al., (1994),  J. Geophys. Res., $\bf 99$, 11251.


\noindent Eilek,J.A., \& Hughes, P.A., 1991, in $Beams\ and\ Jets\ in\
Astrophysics,$ ed. P.A. Hughes,  (New York: Cambridge Univ. Press).


\noindent Feldman,W.C. et al., 1990, J. Geophys. Res., $\bf 90$, 233.





\noindent Jones,F.C., \& Ellison,D.C., 1991, Space Sci. Rev., $\bf
58$, 259.

\noindent Kantrowitz, A., \& Petschek,H.E., 1966  in $Plasma\ Physics\
in \ Theory\ and\ Application$
edited by W. B. Kunkel, (New York: McGraw-Hill).


\ni Larosa, T.N. et al., 1996,  ApJ, {\bf 467}, 454.

\ni Lin R.P. \& Hudson, H.S., 1971, Sol. Phys., {\bf 17}, 412.


\ni Masuda, S. et al., 1994, Nature, {\bf 371}, 495.


\noindent Melrose, D.B., 1986, $Instabilities\ in\ Space\ and\
Laboratory\ Plasmas$ (Cambridge: Cambridge Univ. Press).


\ni Pallavicini, R., Phil. Trans. R. Soc., 1991, A{\bf 336}, 389.

\ni Pneuman, G.W., 1981, in {\it Solar Flare Magnetohydrodynamics},
ed. E.R. Priest, (New York: Gordon \& Breach).


\noindent Omidi, N. \& Winske, D., 1994, 
J. Geophys. Res., $\bf 97$, 14801.

\ni  Petschek, H.E., 1964, AAS-NASA Symp on Phys. of Solar Flares, NASA SP-50, 425.


\ni Priest, E.R. \& Forbes, T.G., 1986, J. Geophys.Res {\bf 91}, 5579.

\ni Forbes, T.G. \& Priest, E.R., 1987, Rev. Geophys. {\bf 25}, 1583.

\noindent Parker, E.N., $Cosmical\ Magnetic\ Fields$, 1979, 
(New York:  Oxford University Press).



\ni Sonnerup, B.U.${\ddot {\rm O}}$, 1970, J. Plasma Phys., {\bf 4} 161.

\ni Sturrock, P.A., 1966, Nature {\bf 211}, 695.

\ni Tsuneta S. et al., 1992, PASJ, {\bf 44}, L63.

\ni Tsuneta, S., 1996a, ApJ, {\bf 456}, 840.

\ni Tsuneta, S., 1996b, in {\it Observations of Magnetic Reconnection
in the Solar Atmosphere}, ed. B. Bentley and J. Mariska, PASP conf., in press.



\vfill
\eject

\centerline{\bf Figure Captions}
\doublespace

{\bf Figure 1}

\noindent Canonical solar flare structure.  The hard X-ray sites
shown are characteristic of smaller CI flares
rather than LDE flares but the overall flare structure
is the same.  The X-point lies within the dissipation region at
the top	between the merging inflows.
Only the downward outflow from the dissipation region is shown
but there is also a vertical outflow.

{\bf Figure 2}

\noindent Schematic of the reconnection region 
flows and fields in the switch off shock case
analyzed in the text. The dotted line is a possible
outflow fast shock.

\end